\documentclass[%
onecolumn,%
oneside,%
floats,%
aps,%
prd,%
nobibnotes,%
nofootinbib,%
amsmath,%
amssymb,%
amsfonts,%
amscd,%
superscriptaddress,%
eqsecnum%
]{revtex4}

\usepackage[utf8]{inputenc}
\usepackage{graphicx, array, dcolumn}
\usepackage[paperwidth=210mm, paperheight=297mm, centering, hmargin=1.8cm, vmargin=2.5cm]{geometry}

\begin{document}

\title{Separated matter and antimatter domains\\ with vanishing domain
  walls}

\author{A.D.\,Dolgov}
\email{dolgov@fe.infn.it} \affiliation{Physics Department and Laboratory of
Cosmology and Elementary Particle Physics, Novosibirsk State University, Pirogova
st.\ 2, Novosibirsk, 630090 Russia} \affiliation{Dipartimento di Fisica e Scienze
della Terra, Universit\`a degli Studi di Ferrara,\\ Via Saragat 1, Ferrara, 44122
Italy}

\author{S.I.\,Godunov}
\email{sgodunov@itep.ru} \affiliation{Physics Department and Laboratory of Cosmology
and Elementary Particle Physics, Novosibirsk State University, Pirogova st.\ 2,
Novosibirsk, 630090 Russia} \affiliation{Theory Division, Institute for Theoretical
and Experimental Physics, Bolshaya Cheremushkinskaya st.\ 25, Moscow, 117218 Russia}

\author{A.S.\,Rudenko}
\email{a.s.rudenko@inp.nsk.su} \affiliation{Physics Department and Laboratory of
Cosmology and Elementary Particle Physics, Novosibirsk State University, Pirogova
st.\ 2, Novosibirsk, 630090 Russia} \affiliation{Theory Division, Budker Institute
of Nuclear Physics, akademika Lavrentieva prospect 11, Novosibirsk, 630090 Russia}

\author{I.I.\,Tkachev}
\email{tkachev@ms2.inr.ac.ru} \affiliation{Physics Department and Laboratory of
Cosmology and Elementary Particle Physics, Novosibirsk State University, Pirogova
st.\ 2, Novosibirsk, 630090 Russia} \affiliation{Experimental Physics Department,
Institute for Nuclear Research of the Russian Academy of Sciences, 60th October
Anniversary prospect 7a, Moscow, 117312 Russia}

\begin{abstract}

We present a model of spontaneous (or dynamical) $C$ and $CP$ violation where it is
possible to generate domains of matter and antimatter separated by cosmologically
large distances. Such $C$($CP$) violation existed only in the early universe and
later it disappeared with the only trace of generated baryonic and/or antibaryonic
domains. So the problem of domain walls in this model does not exist. These features
are achieved through a postulated form of interaction between inflaton and a new
scalar field, realizing short time $C$($CP$) violation.

\end{abstract}

\maketitle

\section{Introduction}

Our local cosmological neighborhood is made of baryons, while fraction of
antimatter, presumably of astrophysical origin, is vanishingly small. So the
observations indicate that the universe is 100\% baryo-asymmetric, at least locally.
The Baryon Asymmetry (of the Universe), BAU, cannot be explained in the frameworks
of the Standard Model (SM) of particle physics. Alongside with evidence for dark
matter and dark energy it is considered as unambiguous proof of the existence of new
physics beyond SM.

Many quite different extensions of the SM and various scenarios for the BAU
generation were suggested in the literature, for a review see e.g.
Refs.~\cite{BAU_1, BAU_2, BAU_3, BAU_4, BAU_5}. Typically, consideration is
restricted to the models where the universe is asymmetric globally. This is the
simplest possibility. However, it is not excluded that the real universe may be
globally symmetric. It may consist of domains of matter and antimatter, and if the
domains are sufficiently large and far away, they may escape observational
constraints on matter-antimatter annihilation at the domain boundaries. In the
simplest version of the scenario the distance to the nearest domain of antimatter
should be close to the present day cosmological horizon~\cite{annihilation_constr}.

Corresponding particle physics models, leading to the universe creation with
abundant antimatter domains were suggested and developed in the past. While being
more involved, the models of this type also suffer from the inherent problem -- a
domain wall problem~\cite{DWProblem}. Indeed, BAU can be generated only if $CP$
violation is sufficiently strong, beyond the SM capabilities. In addition, in
globally symmetric universe $CP$ should have different signs in different domains.
This non-trivial pattern of $CP$ violation could be provided by a dedicated physical
field, one way or another. Therefore, unavoidably, domains with different $CP$ phase
would be separated by domain walls with unacceptably high energy density, in
conflict with observations. There is only one way out of this restriction. Namely,
domains with different sign (and possibly strength) of $CP$ violation should exist
only in the universe past and should disappear by now, all together with domain
walls, though their effects in the form of matter and antimatter objects would
survive to the present day.

One class of models where this can be achieved has been suggested in
Refs.~\cite{Kuzmin_1, Kuzmin_2, Kuzmin_3, Kuzmin_4}. The main idea behind is a
possibility of an unusual symmetry behavior at high temperatures. It is well known
that a symmetry, which is broken in vacuum, at high temperatures tends to be
restored. But in general, the situation is not that simple and straightforward. It
is also possible that a symmetry is broken only in a particular range of
temperatures, i.e. it is restored at the highest as well as at the lowest
temperatures, for the particular models and details see~\cite{Kuzmin_1, Kuzmin_2,
Kuzmin_3, Kuzmin_4}. This is just what is needed for a matter-antimatter domain
generation without domain wall problem. However, if a model is based on the unusual
symmetry behavior at high temperatures, then the size of domains will be too small
from the cosmological point of view. Such models are still interesting, because they
could provide a local excess of antimatter, which would be large compared to the
capabilities of astrophysical sources, if an excess~\cite{AntiExcess} turns out to
be real. But cosmologically large and separated domains of antimatter cannot be
created by this mechanism. Cosmology with domains of matter and antimatter was
discussed in the lectures~\cite{fs, ad}, where a list of relevant references can be
found. As argued in Ref.~\cite{larsson}, domain walls could be also eliminated if
the vacua in the model were not exactly degenerate. In this case the higher energy
vacuum would be "swallowed" by the lower energy one if the energy difference is
sufficiently high.

In the present paper we suggest another scenario of unusual symmetry behavior. Now
this happens during inflationary stage in the universe evolution. Domains with
different sign of $CP$ disappear by now also, so the domain wall problem is absent.
However, they appeared during inflation and survived at the baryogenesis epoch,
therefore, cosmologically large domains of matter and antimatter could be created.

\section{Model}
\label{sec:model}

\begin{figure}[h]
\center
\includegraphics[scale=1.0]{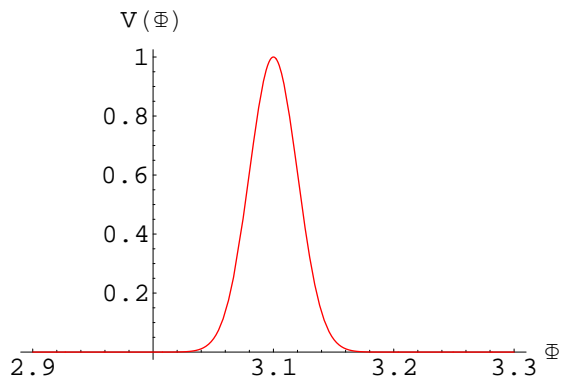}
\caption{\label{fig:V} Function $V(\Phi)$. The parameters are
$\Phi_0=3.1\,m_{Pl}$ and $\Phi_1=0.02\,m_{Pl}$. Field $\Phi$ is
measured in units of $m_{Pl}$ and $V(\Phi)$ is dimensionless.}
\end{figure}

In the suggested model the difference between matter and antimatter is generated by
pseudoscalar field $\chi$ which interacts with inflaton field $\Phi$. We assume the
following Lagrangian:
\begin{equation}
L = L_\Phi + L_\chi + L_{int},
\label{L-tot}
\end{equation}
where
\begin{eqnarray}
L_\Phi &=& \frac{1}{2}(\partial \Phi)^2 - \frac{1}{2} M^2 \Phi^2,
\label{L-Phi} \\
L_\chi &=& \frac{1}{2}(\partial \chi)^2 - \frac{1}{2}m^2 \chi^2 -
\frac{1}{4}\lambda_\chi \chi^4,
\label{L-chi} \\
L_{int} &=& \mu^2  \chi^2 V(\Phi). \label{L-int}
\end{eqnarray}
Here the metric tensor enters into kinetic terms in the usual way, and $M, m,
\lambda_\chi, \mu$ are some constant parameters, with $M, m, \mu$ having dimension
of mass, $\lambda_\chi$ being dimensionless. We do not include the quartic term
$-\lambda_\Phi \Phi^4/4$ in the Lagrangian, though it possibly may lead to some
interesting consequences. This case will be studied elsewhere. The dimensionless
function $V(\Phi)$ is chosen in the way that it is non-zero only when $\Phi$ is
close to some constant value $\Phi_0$. In this paper we choose it as Gaussian
function
\begin{equation}
V(\Phi) = \exp \left[-\frac{(\Phi-\Phi_0)^2}{2\Phi_1^2}\right], \label{V-of-Phi}
\end{equation}
though other forms may be possible. The plot of function $V(\Phi)$ (see
Fig.~\ref{fig:V}) is a bell-shaped curve. Parameters $\Phi_0$ and $\Phi_1$ have
dimension of mass and indicate position of the "bell" center and characteristic
width of the "bell", respectively.

The equations of motion have the form:
\begin{eqnarray}
 \ddot{\Phi}+ 3H \dot{\Phi}+ M^{2} \Phi+
 \mu^{2}\chi^{2} \frac{\Phi-\Phi_{0}}{\Phi_{1}^{2}}V(\Phi) &=& 0,
 \label{ddot-Phi}\\
  \ddot{\chi}+ 3H \dot{\chi}+m^{2}\chi+\lambda_{\chi}\chi^{3} -
  2\mu^{2} \chi V(\Phi) & = & 0,
\label{ddot-chi}
\end{eqnarray}
where $H=\dot a/a$ is the Hubble parameter, $a(t)$ is the cosmological scale factor
which enters into the FLRW metric as
\begin{equation}
ds^2 = dt^2 - a^2(t) \,d{\bf x}^2. \label{ds2}
\end{equation}
It is assumed here that fields $\Phi$ and $\chi$ depend only on time, $\Phi = \Phi
(t)$ and $\chi=\chi(t)$.

The Hubble parameter is expressed through the energy density $\rho$ as
\begin{equation} \label{H}
 H = \sqrt{\frac{8\pi\rho}{3m_{Pl}^{2}}}=
   \sqrt{\frac{8\pi}{3m_{Pl}^{2}}\left(
     \frac{\dot{\Phi}^{2}}{2}
     +\frac{M^{2}\Phi^{2}}{2}
     +\frac{\dot{\chi}^{2}}{2}
     +\frac{m^{2}\chi^{2}}{2}
     +\frac{\lambda_{\chi}\chi^{4}}{4}
     -\mu^{2}\chi^{2}V(\Phi)\right)}\ ,
\end{equation}
where $m_{Pl} \approx 1.2 \cdot 10^{19}$ GeV is the Planck mass.

\begin{figure}[h]
\center
\begin{tabular}{c c c}
\includegraphics[scale=0.7]{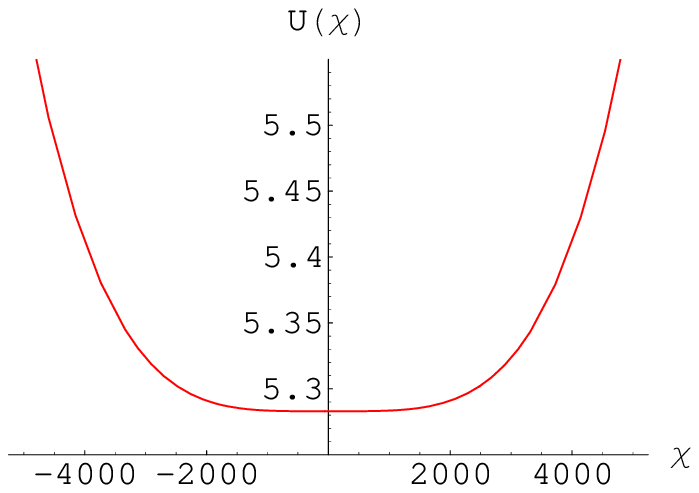} &
\includegraphics[scale=0.7]{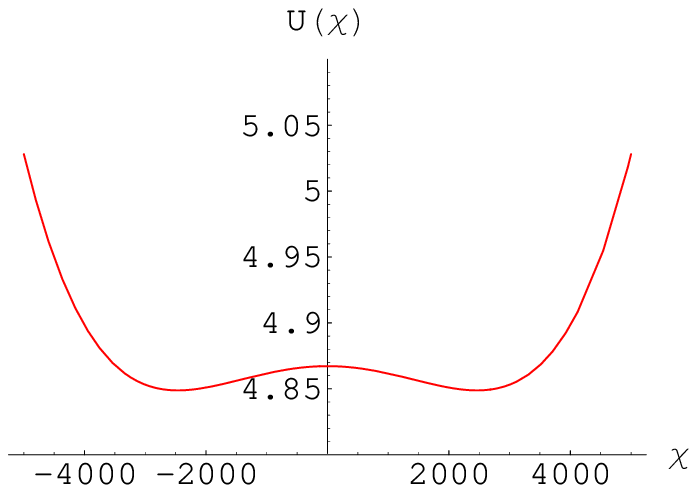} &
\includegraphics[scale=0.7]{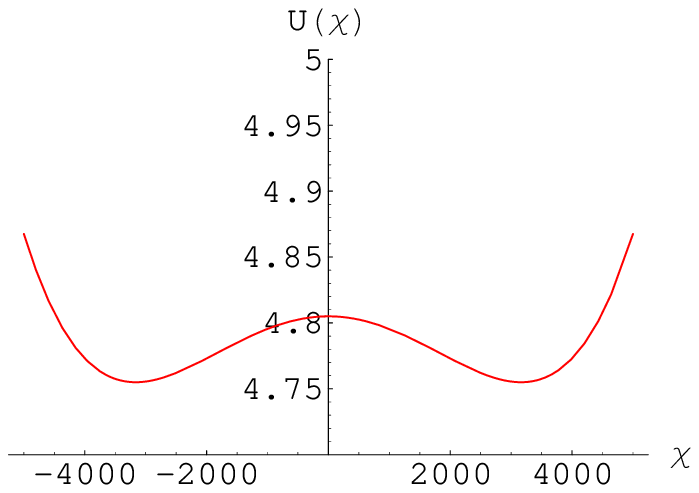}
\end{tabular}
\caption{\label{fig:U-phi-chi} Potential $U(\Phi,\chi)$ at
constant $\Phi$. The left plot corresponds to the moment when the
effective mass of $\chi$ becomes zero, i.e. when $m^2/2-\mu^2
V(\Phi)=0$, therefore
$\Phi=\Phi_0+2\Phi_1\sqrt{\ln{(\sqrt{2}\mu/m)}}$. The middle and
the right plots show the moments when $\Phi=\Phi_0+\Phi_1$ and
$\Phi=\Phi_0$, respectively; the squared effective mass of $\chi$
is negative in these cases. The parameters are
$\Phi_0=3.1\,m_{Pl}$, $\Phi_1=0.02\,m_{Pl}$, $\mu=10^{-4}m_{Pl}$
and $m=10^{-10}m_{Pl}$. Field $\chi$ is measured in units of $M$
and $U(\Phi,\chi)$ is measured in units of $10^{-12}\,m_{Pl}^4$.}
\end{figure}

The interaction introduced above leads to the following scenario. During inflation
the magnitude of the inflaton field $\Phi$ decreases and when it reaches vicinity of
$\Phi_{0}$ \footnote{Of course, we suppose that initial value of inflaton field
$\Phi_{in}$ is larger than $\Phi_{0}$.} two minima appear in the potential
\begin{equation}
U(\Phi,\chi) = \left(\frac{1}{2}m^2-\mu^2 V(\Phi)\right)\chi^2 +
\frac{1}{4}\lambda_\chi \chi^4+\frac{1}{2}M^2\Phi^2 \label{U-phi-chi}
\end{equation}
at constant $\Phi$, so the point $\chi=0$ becomes local maximum (see
Fig.~\ref{fig:U-phi-chi}).

Therefore, in the spatial regions where the field $\chi$ turns out to be positive
(due to fluctuations) it rolls down to the positive minimum $\eta>0$, and in the
regions where $\chi$ is negative it goes to the negative minimum $-\eta<0$. Thus one
can substitute $\chi=\widetilde{\chi}+\eta$ and $\chi=\widetilde{\chi}-\eta$,
respectively, where $\eta$ and $-\eta$ have the meaning of the vacuum expectation
values $<\chi>$. The field $\tilde \chi$ includes quantum fluctuations around new
vacua but in what follows we consider only a classical part of it $\tilde \chi =
\tilde \chi (t)$.

Let us suppose that fermions enter into theory through the
following Lagrangian:
\begin{equation}
L_\psi =
\bar{\psi}(i\hat{\partial}-m_\psi)\psi-g\chi\bar{\psi}i\gamma_5\psi,
\label{L-ferm}
\end{equation}
where $g$ is dimensionless constant. This Lagrangian respects all discrete
symmetries, as the field $\chi$ is $P$-odd and $C$-even. Substituting
$\chi=\widetilde{\chi}\pm\eta$ and carrying out the axial rotation $\psi\rightarrow
{\rm exp}(i\alpha\gamma_5)\,\psi$ with $\alpha=1/2\,{\rm arctan}(\mp\eta g/m_\psi)$
in order to eliminate the term $\bar{\psi}i\gamma_5\psi$, we find
\begin{equation} \label{L-ferm-rot}
L_\psi =
\bar{\psi}(i\hat{\partial}-M_\psi)\psi-g\frac{m_\psi}{M_\psi}\widetilde{\chi}\bar{\psi}i\gamma_5\psi
\mp g^2 \frac{\eta}{M_\psi}\widetilde{\chi}\bar{\psi}\psi,
\end{equation}
where $M_\psi=\sqrt{m_\psi^2+g^2\eta^2}$. The last two terms in (\ref{L-ferm-rot})
behave in opposite ways under $CP$-conjugation, so this Lagrangian violates
$CP$-symmetry. Moreover, the sign of the last term in (\ref{L-ferm-rot}) depends on
the choice of the minimum (positive or negative) where the field $\chi$ has rolled
down. Thus all the space turns out to be divided into domains with opposite signs of
$CP$ violation~\cite{DWProblem}. These domains are separated by domain walls which
should vanish at the present epoch, because the field $\chi$ ultimately tends to the
final minimum at $\chi = 0$ independently of its initial position at $\chi =
\pm\,\eta$, as we see in what follows.

We expect that the distances between domains, as well as domain sizes exponentially
grow with the scale factor $a(t)$, so at the present time these domains are of
cosmological size and they are separated by large distance which prevents the
annihilation at their boundaries. According to Ref.~\cite{annihilation_constr}, a
very large piece of space between the domains, devoid of baryons, would lead to too
large angular fluctuations of the CMB temperature. On the other hand, if the
distance between domains is smaller than the baryon diffusion length, they would be
able to meet successfully their counterparts and to annihilate. These arguments
resulted in the conclusion that the nearest domains should be at the cosmologically
large distance from us, about a few Gpc. On the other hand, the effects of the
cosmological inhomogeneity, not yet studied, could inhibit the baryon diffusion and
would relax the bound on the distance to the nearest antimatter domain.

However, $CP$ violation described by Lagrangian (\ref{L-ferm-rot}) is operative only
when the field $\chi$ sits near the temporary minimum, $\pm\,\eta$, i.e. when the
expression $\chi=\widetilde{\chi}\pm\eta$ is valid, where $\widetilde{\chi}$ are
small quantum fluctuations. In our model this is true only during inflation.
Consequently such $CP$ violation is not efficient for baryogenesis, because the
generated at this period baryon asymmetry would be exponentially inflated away.
Hence successful baryogenesis should take place after the end of inflation. Though
the minima at $\chi = \pm \eta$ disappeared after inflation, still the classical
field $\chi(t)$ remained non-vanishing for a long time after inflation was over. The
classical field $\chi$ slowly tends to zero and if baryogenesis could proceed fast
enough, while $\chi \neq 0$, the $CP$ violation induced by non-zero $\chi (t)$ might
be effective. In these circumstances more efficient $CP$ violation would originate
from the imaginary part of the quark effective mass matrix proportional to $\chi(t)$
and to this end  at least 3 quark families are necessary~\cite{KM}. The only
difference with the standard case is that the contribution to $CP$-odd phase of the
mass matrix is not constant anymore, but slowly changes with time. In more detail
this is described below at the end of section~\ref{sec:production}.

Now let us consider what happens at different stages of the scenario, so we will be
able to put the limits on initial conditions and parameters of the model.

For simplicity we assume, though it is not necessary, that the impact of field
$\chi$ on cosmological expansion at inflationary stage is negligible. To this end we
suppose that the energy density of $\Phi$ dominates over that of $\chi$ and the
energy density of their interaction. Since the field $\Phi$ is not affected by
$\chi$, the evolution of $\Phi$ is described by standard inflation theory, which is
taken here as the usual slow-roll regime of inflation. In this approximation the
second derivative term and the interaction term proportional to $V(\Phi)$ in the
inflaton equation of motion (\ref{ddot-Phi}) can be neglected. Therefore, we find
\begin{equation}
 \dot \Phi = - \frac{M^2 \Phi}{3 H}.
\label{dot-Phi}
\end{equation}
We suppose that the term $M^2\Phi^2/2$ makes the dominant contribution to the energy
density $\rho$, so the Hubble parameter (\ref{H}) is equal to $H=\sqrt{4\pi/3}\,
(M/m_{Pl}) \Phi$, and the equation (\ref{dot-Phi}) can be easily integrated giving
\begin{equation} \label{Phi-of-t}
\Phi(t) = \Phi_{in} - \frac{M m_{Pl}\, t}{2\sqrt{3\pi}} ,
\end{equation}
where $\Phi_{in}$ is an initial value of $\Phi$. In this scenario of inflation the
Hubble parameter gradually decreases, $H\sim \Phi$, and to obtain a sufficiently long
inflation with duration
\begin{equation}
  \int\limits_{0}^{t_{end}}H(t) dt \approx\frac{1}{2}H_{in}t_{end}
  =\frac{2\pi\Phi^2_{in}}{m^2_{Pl}}>70,
\label{H-Delta-t}
\end{equation}
it is sufficient to take $\Phi_{in} \gtrsim 3.3\,m_{Pl}$. In Eq. (\ref{H-Delta-t})
$H_{in}= \sqrt{4\pi/3}\,(M/m_{Pl}) \Phi_{in}$ is the initial value of the Hubble
parameter and $t_{end} = 2\sqrt{3\pi}\,\Phi_{in} /(M m_{Pl})$ is the time moment
when $\Phi = 0$, see (\ref{Phi-of-t}). To be more precise, inflation ends when the
Hubble parameter becomes approximately equal to the mass of the inflaton,
$H_{fin}\sim M$, because at this moment the exponential expansion terminated and the
inflaton field started to oscillate around zero with frequency $M$. Correspondingly
the exponential expansion turned into the matter dominated one, $a \sim t^{2/3}$. So
the final amplitude of $\Phi$ should be taken as $\Phi_{fin} =
\sqrt{3/(4\pi)}\,m_{Pl}$. Correspondingly the final time would be slightly changed.

In order to avoid too large density perturbations one should choose the value of the
inflaton mass $M$ in the range $10^{-7}\lesssim M/m_{Pl}\lesssim 10^{-6}$; accordingly  we
take $M=10^{-6}\,m_{Pl}$.

To arrange the desired scenario we should set the value of
$\Phi_{0}$ still at inflationary stage such that
the distances between different domains and the domain sizes
exponentially expanded up to cosmologically large scales.

We can make a naive estimate of the necessary duration of inflation after the
inflaton field reached value $ \Phi = \Phi_0$. Suppose that at the present time the
size of a domain is about 10 Mpc. If the characteristic scale of the initial
energy/temperature is of the order of $T$, then it is natural to expect the domain
size to be of the order of $ l\sim 1/T$. Due to regular cosmological expansion it
would increase to $z =(1/T)(T/3K)\sim 0.1$ cm. So during inflation the domains
should be expanded by a factor of (10 Mpc)/(0.1 cm) $\sim 3\cdot 10^{26}$.
Therefore, we require that after $\Phi$ passes $\Phi_{0}$ the inflation should last
at least 60 e-foldings, $N_{e}={\rm ln}\, (3\cdot 10^{26}) \simeq 60$,
\begin{eqnarray}
  \label{eq:Phi0}
  \int\limits_{t_{0}}^{t_{end}}H(t)dt
  =\frac{2\pi\Phi_{0}^{2}}{m_{Pl}^{2}}\gtrsim N_{e}\simeq 60,\\
  \frac{\Phi_{0}}{m_{Pl}}=\sqrt{\frac{N_{e}}{2\pi}}\approx 3.1,
\end{eqnarray}
where $t_{0}$ is the time when $\Phi=\Phi_{0}$. However, since the process initiated
at inflationary stage, when temperature in classical sense did not exist, the
characteristic scale of the seed of the domain should be of the order of the inverse
Hubble parameter, $l \sim 1/H \sim 1/M$. So the obtained above result would be
shifted to $N \approx 60 + \ln (M/T_{heat})$, where $T_{heat}$ is the universe
temperature after inflation. Since typically $T_{heat} \approx (10^{-5} \div
10^{-6})\, m_{Pl}$ the estimate presented above remains practically unchanged.
\\

Depending upon the value of $\Phi$, there are three different regimes of the
evolution of the field $\chi$:
\begin{enumerate}
 \item  Initial stage: $\Phi>\Phi_{0}$ and $\left(\Phi-\Phi_{0}\right)\gg\Phi_{1}$.

   The mass of $\chi$ is supposed to be small in comparison with the
   expansion rate, $m<H$, so field $\chi$ sits near the minimum
   of its potential at $\chi = 0$ (see left plot in Fig.~\ref{fig:U-phi-chi}), with the
   dispersion of the order of $\langle \chi^2 \rangle \sim {\rm min} \{H^4/m^2,H^2/\sqrt{\lambda_\chi}\}$ (see e.g.~\cite{Gorbunov}).

 \item Second stage: $\left|\Phi-\Phi_{0}\right|\lesssim \Phi_{1}$.

   At this stage the squared effective mass of $\chi$ becomes negative:
   $m^2-2\mu^2V(\Phi)\approx -2\mu^2V(\Phi)<0$, if $m$ is small enough.
   The minima of the potential (\ref{U-phi-chi}) move from
   zero as $\eta(t)$ or $-\eta(t)$, where
   \begin{equation}
    \eta^2(t) = \frac{1}{\lambda_\chi}\left(2\mu^2V(\Phi)-m^2\right).
    \label{chi-min-of-t}
   \end{equation}

To have equal amplitudes of both positive and negative $CP$ violation we need to
impose the condition that $\chi$ reaches one or other minimum $\chi=\pm\,\eta(t)$
and stays there (approximately). This minimum moves exponentially fast and $\chi$
would follow it e.g. if $\mu\gg H$, since Eq.~(\ref{ddot-chi}) becomes then
$\ddot{\chi}-2\mu^{2}\chi V(\Phi)=0$,
   therefore $\chi(t)$ grows roughly as exponent, $\chi \propto \exp (\mu t)$.
   For $M=10^{-6}m_{Pl},~\Phi_{0}=3.1\,m_{Pl}$ we get $\mu\gg H=\sqrt{4\pi/3}\,M/m_{Pl}\,\Phi \sim 6\cdot
   10^{-6}m_{Pl}$.

   But even when $\chi$ grows exponentially there still should be
   enough time for $\chi$ to reach the minimum $\eta(t)$. If the inflaton $\Phi$ goes from $\Phi_{0}+2\Phi_{1}$ to
   $\Phi_{0}-2\Phi_{1}$ during the interval $\tau$, we should require that
   \begin{eqnarray}
     \label{eq:mutau}
     \mu\tau=\mu\frac{8\sqrt{3\pi}\Phi_{1}}{Mm_{Pl}}\gtrsim\ln\frac{\eta_{max}}{\chi_{in}}
     =\frac{1}{2}\ln\frac{2\mu^{2}}{\lambda_{\chi}\chi_{in}^{2}},\\
     \mu\gtrsim\frac{Mm_{Pl}}{16\sqrt{3\pi}\Phi_{1}}\ln\frac{2\mu^{2}}{\lambda_{\chi}\chi_{in}^{2}},
   \end{eqnarray}
   where $\eta_{max}\equiv 2\mu^{2}/\lambda_{\chi}$ and $\chi_{in}$ is
   initial value of $\chi$.

   According to this scenario
   $\chi^{2}\sim\eta_{max}^{2}\equiv 2\mu^{2}/\lambda_{\chi}$ is the
   largest value of $\chi^2$. So to be sure that the inflaton field $\Phi$
   always gives the main contribution to the energy density, we should
   impose the conditions:
   \begin{eqnarray}
     \label{eq:Phi>chi}
     M^{2}\Phi_{0}^{2}\gg\mu^{2}\left.\chi^{2}\right|_{\Phi=\Phi_{0}}
    \sim\frac{\mu^{4}}{\lambda_{\chi}},\\
     \mu^{4}\ll M^{2}\Phi_{0}^{2}\lambda_{\chi}.
   \end{eqnarray}
   For $M=10^{-6}m_{Pl},~\Phi_{0}=3.1\,m_{Pl}$ we get $\mu\ll 1.8\cdot
   10^{-3}m_{Pl} \sqrt[4]{\lambda_{\chi}}$.

   After $\Phi$ passed $\Phi_{0}$, $\eta(t)$ started to
   decrease exponentially and turned zero when
   $\Phi=\Phi_0-2\Phi_1\sqrt{\ln{(\sqrt{2}\mu/m)}}$, so the potential
   (\ref{U-phi-chi}) started to have
   only one minimum at $\chi=0$ again. The
   field $\chi$ also began to decrease rapidly until $V(\Phi)$ became small
   in comparison with other terms in Eq.~(\ref{ddot-chi}).

\item Final stage: $\Phi<\Phi_{0}$ and
     $\left(\Phi_{0}-\Phi\right)\gg\Phi_{1}$.

   When the interaction term $\mu^2 \chi^2 V(\Phi)$ vanishes, the evolution of field $\chi$
   would be determined by the terms with $\lambda_{\chi}$ and $m$, see Eq.~(\ref{ddot-chi}).
   If $\lambda_{\chi}$ is not small enough and $\chi$ is still large
   then the term $\lambda_{\chi}\chi^{3}$ dominates and the equation of motion turns into:
   \begin{eqnarray}
     \label{eq:Hlambda}
     3H\dot\chi+\lambda_{\chi}\chi^{3}=0,
 \end{eqnarray}
 which has the solution:
 \begin{eqnarray}
     \chi=\sqrt{\frac{3H}{2\lambda_{\chi}}}\frac{1}{\sqrt{t-C}}\,,
   \label{chi-sol}
   \end{eqnarray}
   where $C$ is a constant of integration. We see that in this
   regime $\chi$ decreases as a power of $t$.

   When $\chi$ becomes quite small, Eq.~(\ref{ddot-chi}) turns into
   $\ddot\chi+3H\dot\chi+m^2\chi=0$, so the field $\chi$ slowly oscillates and decreases due to redshift related to cosmological
   expansion.
\end{enumerate}

An important issue for the considered model is the character of the domain wall
expansion during inflation, when the field $\chi$ is situated near the wall center
at $\chi=0$. This problem was studied in Refs.~\cite{linde, vilenkin_1, vilenkin_2}.
It was shown that for narrow wall, when its width was shorter than the inverse
Hubble parameter, the wall size remained narrow as in the flat space-time. However,
if the width is large, then the cosmological expansion would stretch the wall
exponentially and moreover inflation would proceed inside the wall. In our case, as
we will see in the next section, the domain wall does expand but with the chosen
model parameters there is no big difference between the expansion of space near the
wall and far from it during the period of wall existence. An essential difference
between our model and those considered in the literature is that in our case the
wall existed only for relatively short time.

\section{Numerical calculations}
\label{sec:num}

In order to check that the described scenario is indeed operative we performed the
numerical calculations in the homogeneous case with the following set of parameters
(all dimensional
    values are given in the Planck mass, $m_{Pl}$, units):
\begin{equation}
  \label{eq:parameters}
  \Phi_{in}=4,\ \Phi_{0}=3.1,\ \Phi_{1}=0.02,\ M=10^{-6},\ \chi_{in}=10^{-6},\ m=10^{-10},\ \lambda_{\chi}=2\cdot10^{-3},\ \mu=10^{-4}.
\end{equation}

\begin{figure}[h]
\center
\begin{tabular}{c c}
\includegraphics[scale=1.0]{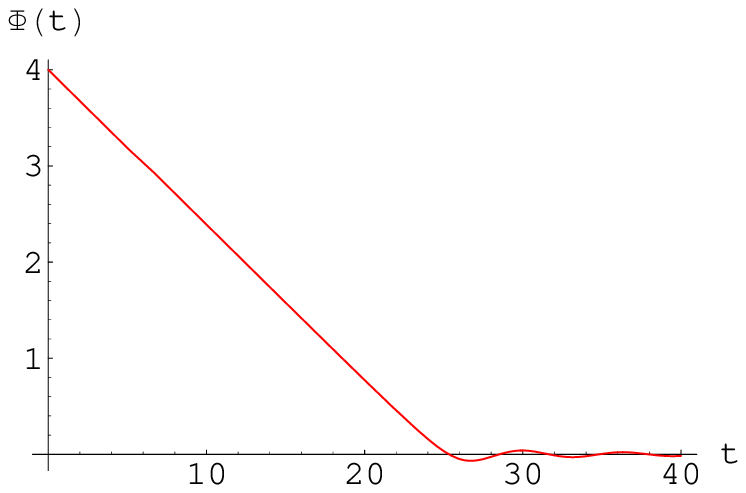} &
\includegraphics[scale=1.0]{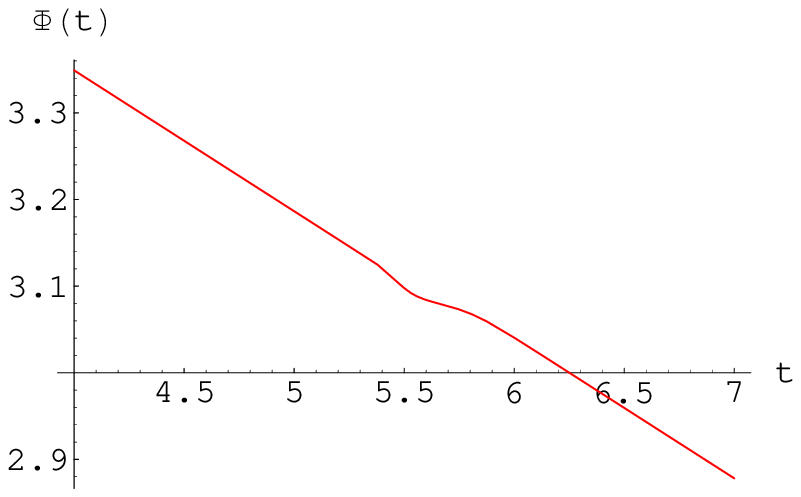}
\end{tabular}
\caption{\label{fig:Phi} Time evolution of the inflaton field $\Phi(t)$ for the
parameters (\ref{eq:parameters}). Time $t$ is measured in units of $M^{-1}$ and
$\Phi$ is measured in units of $m_{Pl}$. In the left plot the slow-roll regime
during inflation and the beginning of the inflaton oscillations are presented. The
right plot shows the evolution of $\Phi$ in more detail near the point
$\Phi=\Phi_{0}=3.1\,m_{Pl}$. The interaction of $\Phi$ with the field $\chi$ has a
negligible effect on the inflation for the chosen parameters, so $\Phi(t)$ only
slightly deviates from the straight line around $\Phi=\Phi_{0}$ (see the right
plot).}
\end{figure}

\begin{figure}[h]
\center
\begin{tabular}{c c c}
\includegraphics[scale=0.7]{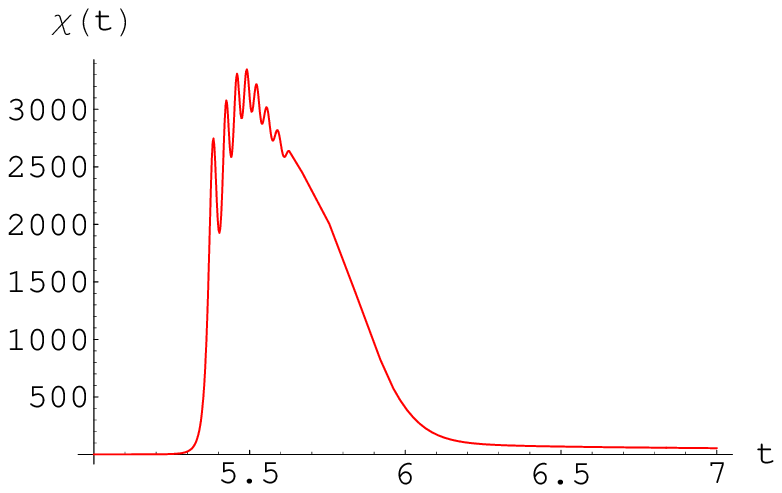} &
\includegraphics[scale=0.7]{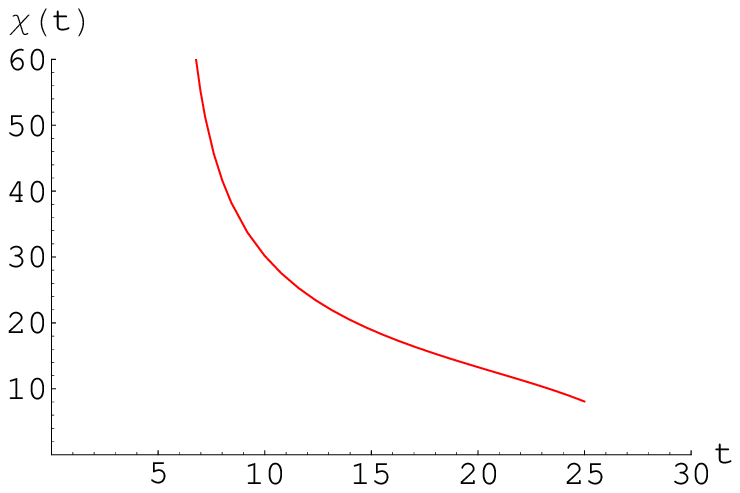} &
\includegraphics[scale=0.7]{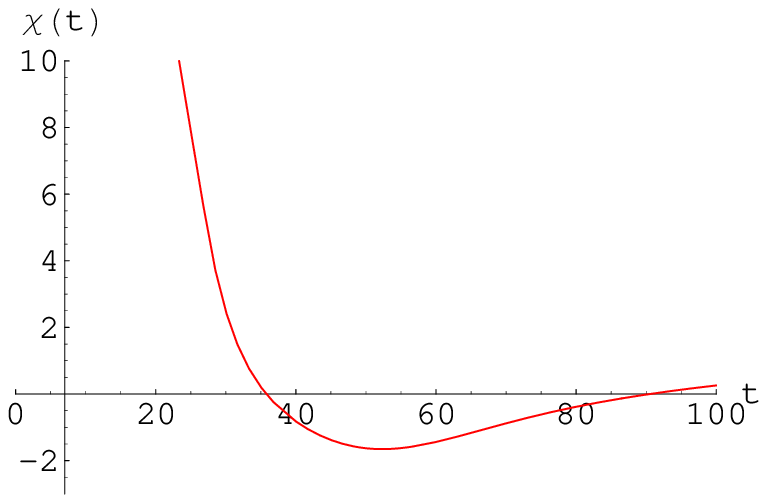}
\end{tabular}
\caption{\label{fig:chi} Time dependence of the field $\chi(t)$ for the parameters
(\ref{eq:parameters}). Time $t$ is measured in units of $M^{-1}$ and $\chi$ is
measured in units of $M$. All the three plots depict the  evolution of $\chi$ during
different time intervals. The left plot shows how the field $\chi$ rolls down to the
minimum of the potential $\eta(t)>0$, oscillates there and then goes back to the
newly formed minimum $\chi=0$. In the middle and right plots one can see that at the
end of inflation, $t\approx 25\,M^{-1}$, the field $\chi$ remains quite large.
Eventually $\chi$ crosses zero and slowly oscillates around $\chi=0$ (see the right
plot).}
\end{figure}

In Figs.~\ref{fig:Phi} and \ref{fig:chi} the results of numerical simulation are
shown. One can see that the interaction of the inflaton field $\Phi$ with the field
$\chi$ does not break the standard inflation, $\Phi(t)$ only slightly deviates from
straight line around $\Phi=\Phi_{0}=3.1\,m_{Pl}$.

In Fig.~\ref{fig:chi} one can see how the field $\chi$ rolls down to the minimum of
the potential $\eta(t)>0$ and oscillates there. Then it goes back to the old minimum
at $\chi =0$, but even at the end of inflation $\chi$ remains an order of magnitude
larger than it was at the beginning ($\chi \sim 10\,M$ at $t\approx 25\,M^{-1}$,
while $\chi_{in}=1\,M$). At the moment $t\approx 35\,M^{-1}$ the field $\chi$
crosses zero and starts slow oscillations with very low frequency, which could be
much lower than the characteristic rate of baryogenesis.

So we see that parameters (\ref{eq:parameters}) are chosen in a proper way. In this
model the influence of $\chi$ on the process of inflation is insignificant, thus the
domain size grows exponentially. However, as we have mentioned at the end of the
previous section, it is also important to know how the regions where $\chi=0$ expand
during inflation. The hypothetical situation is possible when the areas where field
is zero expand exponentially faster than the areas where field is non-zero. For
example, the model where inflation never stops near the domain walls, i.e. near
$\phi=0$, was suggested in \cite{linde}. Also the idea of never ending topological
inflation where $\phi$ is forced to stay near the maximum of its potential $\phi=0$
was proposed in \cite{vilenkin_1, vilenkin_2}. However, our situation is quite
different. Numerical calculation shows that if $\chi=0$ the Hubble parameter remains
almost the same as in the case of non-zero $\chi$, and the size of region where
$\chi=0$ is approximately two times less than the domain size. Therefore, our model
predicts the domains of cosmological size with the distances between them of the
same order of magnitude.

\section{Production of heavy particles after inflation and baryogenesis} \label{sec:production}

As it is commonly known, the stage of inflation is followed by the stage of
(re)heating, during which the very heavy $X$ particles can be produced through decay
of inflaton field (see e.g. book~\cite{Gorbunov}) or from vacuum fluctuations in
gravitational field~\cite{KT}.

Let us consider the situation when the inflaton field $\Phi$ interacts with $X$
particles (for example, scalar bosons) through the coupling $g^2_X\Phi^2 X^2$. As is
shown in Refs.~\cite{brandenberger, kls_1, kls_2}, the particle production can be
strongly enhanced due to parametric resonance. For an unsuppressed production it is
essential that the resonance is broad. The corresponding condition has the form:
\begin{equation}
  \label{broadres}
  g_X \Phi_e \gg M.
\end{equation}
Here $M \simeq 10^{-6} m_{Pl}$ is the inflaton mass, and $\Phi_e \simeq
m_{Pl}/\sqrt{4\pi}$ is the magnitude of the inflaton field at the end of inflation
in our model. One can consider $\Phi_e$ also as the initial amplitude of the
inflaton oscillations. The condition (\ref{broadres}) is true for the wide range of
coupling constant, $g_X \gg 4\cdot 10^{-6}$.

Because of resonance (\ref{broadres}), very heavy $X$-bosons with masses even of GUT
scale $m_X \sim (M m_{Pl})^{1/2} \sim 10^{16}$ GeV can be produced under quite
reasonable assumption $g_X \sim 1$ \cite{Gorbunov}. Moreover in this case of
comparatively large coupling constant $g_X \sim 1$ the decay of the inflaton field
occurs very rapidly, during only one or few oscillations.

Let us assume that the produced $X$-bosons in turn decay into
fermions, for example, into quark-quark and antiquark-antilepton
pairs, $X \rightarrow qq$ and $X \rightarrow \bar{q}\bar{l}$,
respectively. If the corresponding coupling constants are large
enough, the $X$-bosons decay very quickly. Therefore, the field
$\chi(t)$ may remain non-zero yet to the moment when $X$-bosons
have been completely decayed, indeed $\chi \sim 2\,M \sim 10^{13}$
GeV at $t=30\,M^{-1}$ (see Fig.~\ref{fig:chi}).

The field $\chi$ is real and pseudoscalar and interacts with the
produced fermions as (cf. (\ref{L-ferm}))
\begin{equation}
L_{\chi\psi\psi} = g_{kl}\chi\bar{\psi}^k i\gamma_5\psi^l =
g_{kl}\chi (\bar{\psi}^k_R i\gamma_5\psi^l_L+\bar{\psi}^k_L
i\gamma_5\psi^l_R) = i g_{kl}\chi (\bar{\psi}^k_L \psi^l_R -
\bar{\psi}^k_R \psi^l_L), \label{L-ferm-chi}
\end{equation}
where $k$ and $l$ denote the fermion flavor (sum over repetitive indices is
assumed), $\psi_R=((1+\gamma_5)/2)\psi$, $\psi_L=((1-\gamma_5)/2)\psi$. From the
hermicity of the interaction it follows that $g_{kl}$ is real for $k=l$, and
$g_{kl}=g^*_{lk}$ for $k \neq l$. Since $\chi$ is supposed to be electrically
neutral, it interacts with quarks with the same electric charge, so $k$ and $l$
either run over $u,\,c,\,t$ or $d,\,s,\,b$ and there are no cross terms.

The Lagrangian of free fermions can be written as follows
\begin{equation}
L_{free} = \bar{\psi}^k i\hat{\partial} \psi^k - m_{\psi kl}
\bar{\psi}^k\psi^l = \bar{\psi}^k_R i\hat{\partial}\psi^k_R +
\bar{\psi}^k_L i\hat{\partial}\psi^k_L - m_{\psi kl}
(\bar{\psi}^k_R \psi^l_L + \bar{\psi}^k_L \psi^l_R).
\label{L-ferm-free}
\end{equation}

Therefore, the sum of $L_{free}$ and $L_{\chi\psi\psi}$ can be
presented in matrix form as
\begin{equation}
L_{free}+L_{\chi\psi\psi} = \bar{\psi}_R i\hat{\partial}\psi_R + \bar{\psi}_L
i\hat{\partial}\psi_L - (\bar{\psi}_R M_{\psi}\psi_L + \bar{\psi}_L M_{\psi}^\dag
\psi_R), \label{L-ferm-free-chi}
\end{equation}
here $M_{\psi}=m_{\psi}+ig\chi$ is a non-Hermitian matrix, whereas
$m_{\psi}$ and $g$ are Hermitian ones.

Using simultaneously two unitary transformations $\psi_R \rightarrow
\psi'_R=U_R\psi_R$ and $\psi_L \rightarrow \psi'_L=U_L\psi_L$ one can always
diagonalize the mass matrix in (\ref{L-ferm-free-chi}). Therefore, the interaction
of fermions with pseudoscalar field $\chi$ can be "rotated away" (the elements of
transformation matrices $U_R$ and $U_L$ must depend on the magnitude of $\chi$ in
this case), and the Lagrangian (\ref{L-ferm-free-chi}) takes the simple form
\begin{equation}
L'_{free} = \bar{\psi}^a i\hat{\partial} \psi^a - m'_{\psi ab} \bar{\psi}^a\psi^b,
\label{L-ferm-free-1}
\end{equation}
where $\psi^a$ and $\psi^b$ are the mass eigenstates and correspondingly $m'_{\psi}$
is diagonal matrix with real diagonal elements.

Quite similarly one can transform into mass term the interaction of fermions with
any scalar $g_{Skl} S \bar{\psi}^k \psi^l$ and pseudoscalar $g_{Pkl} P \bar{\psi}^k
i\gamma_5 \psi^l$ fields . However, the interaction of fermions with vector (gauge)
boson $X$ remains the same under these transformations:
\begin{equation} \label{vecint}
g_{Rkl} X_{\mu} \bar{\psi}_R^k \gamma^{\mu} \psi_R^l + g_{Lkl} X_{\mu}
\bar{\psi}_L^k \gamma^{\mu} \psi_L^l \rightarrow g'_{Rab} X_{\mu} \bar{\psi}^a_R
\gamma^{\mu} \psi^b_R + g'_{Lab} X_{\mu} \bar{\psi}^a_L \gamma^{\mu} \psi^b_L,
\end{equation}
here $g'_R = U_R g_R U^\dag_R$ and $g'_L = U_L g_L U^\dag_L$ are matrices of
coupling constants in mass eigenstate basis, so $\psi^a$ describes $a$-th sort of
fermion with definite mass. The constants $g'_{ab}$ are complex in general case, and
if there are at least three species of fermions, one cannot rotate away
simultaneously all phases in complex matrices $g'_{R,L}$ \cite{KM}. The complexity
of the coupling constants means that $CP$ is violated in the $X$-boson decays
\cite{NW}. The magnitude of this $CP$ violation depends on the value of the field
$\chi$ through the matrices $U_{R,L}$ and coupling constants $g'_{R,L}$. Since
$\chi$ is essentially non-zero after the end of inflation and during baryogenesis,
the $CP$-odd effects can be large enough.

We assume also that gauge interactions involve fermions with certain chirality, see
(\ref{vecint}), and thus these interactions break $C$-invariance.

$C$ and $CP$ violation is one of the necessary Sakharov conditions of generation of
baryon asymmetry~\cite{sakharov}. The another one is baryon number violation, so one
needs to assume also that in the decays of $X$-bosons the baryon number is not
conserved. Let $\delta$ be the baryon asymmetry generated in the decay of one
$X$-boson. Then it can be easily demonstrated \cite{Gorbunov} that the ratio of the
baryon number density to the entropy density is estimated as
\begin{equation}
\Delta_B = \frac{n_B - n_{\bar{B}}}{s} \sim \delta \frac{h}{g_X}
\left(\frac{m_{th}}{m_{Pl}}\right)^{1/2}, \label{Bar-asym}
\end{equation}
where $h$ and $g_X$ are typical coupling constants of $X$-boson with fermions and
inflaton, respectively, $m_{th}$ is mass scale of the theory. It is quite reasonable
to believe that $h/g_X \sim 1$ and $m_{th} \sim M \sim 10^{-6}\,m_{Pl}$, so one has
$\Delta_B \sim 10^{-3} \delta$. Thus, to get observed value $\Delta_B \simeq
0.86\cdot 10^{-10}$ it is sufficient to have only $\delta \sim 10^{-7}$. Such small
$\delta$ seems to be easily produced in the decay of $X$-boson. Therefore, the
observed baryon asymmetry of the universe can be generated in the decays of
$X$-bosons, without fine tuning of parameters of the theory.

We have to stress that the baryogenesis should proceed after the end of inflation
(as it has been already mentioned above). Otherwise the baryon asymmetry would be
strongly diluted by the universe (re)heating. When inflation was over, the field
$\chi$ evolved down to $\chi = 0$ and so $\chi$ was not equal to constant $\eta$ but
had considerably smaller and time dependent value. However, since at this stage
$\chi(t)$ changed very slowly one can repeat the above arguments with adiabatically
evolving $\chi (t)$. If baryogenesis proceeded faster than $\chi$ evolved, then the
effective amplitude of $CP$ violation might be considered as a constant and the
corresponding mass matrix could be diagonalized in the same way as it has been done
above. The phase of the mass matrix would be unsuppressed if $g\chi(t)$ was close to
the value of the bare quark mass. Evidently in this case the imaginary part of the
mass matrix would be of the same order of magnitude as the real part.

\section{Conclusions}

We have presented here a model of baryogenesis which may lead to baryo-symmetric
universe with cosmologically large domains of matter and antimatter, avoiding the
domain wall problem. The model satisfies three Sakharov criteria for successful
baryogenesis: non-conservation of the baryon number, deviation from thermal
equilibrium, and $C$ and $CP$ violation. However, the latter is different from the
normally exploited one. Breaking of charge symmetry is induced by a non-zero
amplitude of a scalar field $\chi$, which slowly relaxed down to equilibrium, much
slower than the process of baryogenesis goes on.  In classification of different
types of $CP$ violation which might be operative in cosmology this type is called
dynamical one~\cite{AD-Varenna}. Later, after the baryon asymmetry was developed,
$\chi$ evolved down to the equilibrium point $\chi = 0$ and thus domain walls
disappeared rather early in the universe.

Inflation is an essential ingredient of the scenario. A coupling to the inflaton field was introduced on purpose
to generate a non-zero value of $\chi$ and to keep it non-zero during baryogenesis.

The model allows for successful baryo-symmetric cosmology without
yet being drawn into astronomical controversy.

\section*{Acknowledgement}
We acknowledge support of the Russian Federation Government Grant
No.\,11.G34.31.0047. S.\,G. is partially supported under the
grants RFBR No.\,14-02-00995 and NSh-3830.2014.2. S.\,G. is also
supported by MK-4234.2015.2. and by Dynasty Foundation.

\end{document}